\documentclass[journal]{IEEEtran}
\usepackage{amssymb}
\usepackage{amsfonts}
\usepackage{epsfig,graphics,subfigure,amsthm,psfrag,amsmath}

{\newtheorem{Thm}{Theorem}} {\theoremstyle{plain}
\newtheorem{Lem}{Lemma}}
\newtheorem{Cor}{Corollory}

\newtheorem{Proof}{Proof}
\newtheorem{Problem}{Problem}
\small\normalsize

\begin{document}
%
\title{Design and Analysis of Multi-User SDMA Systems with Noisy Limited CSIT Feedback}
%
%
\author{Tianyu Wu,   Vincent K. N.  Lau
\\wuty@ust.hk,   eeknlau@ee.ust.hk
\\Dept of EEE, Hong Kong University of Science and Technology
\thanks{%
This paper is supported by RGC funding 615606.}}
%
%
%
\markboth{IEEE Transactions on Wireless Communications, Vol. XX, No.
XX, Month 2010}{T. Wu \MakeLowercase{\textit{et al.}}: Design and Analysis of Multi-User SDMA Systems with Noisy Limited CSIT Feedback}
%



\maketitle


%
\IEEEpeerreviewmaketitle

\begin{abstract}
In this paper, we consider spatial-division multiple-access (SDMA)
systems with one base station with multiple antennae and a number of
single antenna mobiles under noisy limited CSIT feedback. We
propose a robust noisy limited feedback design for SDMA systems.
The solution consists of a real-time robust SDMA precoding, user
selection and rate adaptation as well as an offline feedback index
assignment algorithm. The index assignment problem is cast into a
Traveling Sales Man problem (TSP). Based on the specific structure
of the feedback constellation and the precoder, we derive a low
complex but asymptotically optimal solution. Simulation results show
that the proposed framework has significant goodput gain compared to
the traditional naive designs under noisy limited feedback channel.
Furthermore, we show that the average system goodput scales as
$\mathcal{O}(\frac{n_T(1-\epsilon)}{n_T-1}(C_{fb}-\log_2(N_n)))$ and
$\mathcal{O}(n_T\cdot\log_2P)$ in the interference limited regime
($C_{fb} < (n_T-1)\log_2P+\log_2N_n$) and noise-limited regime
respectively. Hence, despite the noisy feedback channel, the average
SDMA system goodput grows with the number of feedback bits in the
interference limited regime while in noise limited regime increases
linearly with the number of transmit antenna and the forward channel
SNR ($\log_2P$).
\end{abstract}

\section{Introduction}
\label{sect:intro} It is widely known that spatial-division
multiple-access(SDMA) is an important technique to enhance the
throughput of multi-user wireless systems due to spatial
multiplexing. However, SDMA system
requires channel state information at transmitter(CSIT). In FDD
systems, only a limited number of bits(e.g. 6 bits for WiMAX
\cite{Wimax:05}) can be allocated to carry the CSIT feedback, namely
the {\em limited feedback}. In \cite{Wang:07}, the authors consider multiuser MISO system with a
limited total feedback bits constraint and proposed a codebook
design algorithm and a CSIT decomposition algorithm. The authors of
\cite{Kaibin:07} analyzed the asymptotic performance of a per user
unitary and rate control design. Other works like
\cite{Murch:04}-\cite{Goldsmith:06} studied transmit beamforming
using different criteria and methods. However, in all these works,
the focus was to study the quantization effects on the CSIT under
noiseless feedback\footnote{The CSIT feedback index is always
received correctly at the transmitter.}. In practice, the CSIT
feedback may not be error-free due to the feedback channel noise.
Unlike the forward channel where heavy error correction coding can
be applied to the time-interleaved payload, the limited CSIT
feedback has to be received at the transmitter with minimum latency
and hence, time interleaving is not possible. Furthermore, in most
systems, the number of bits available for feedback is very limited
(such as 6 bits in WiMAX) and hence, it will be more effective to
utilize all the limited bits to carry the CSIT rather than wasting
some bits to protect the CSIT feedback. The issue of noisy feedback is considered in \cite{Murthy:08, Zhang:08,Lim:09, Housfater:09}.
For example, in \cite{Murthy:08, Zhang:08,Lim:09}, the authors analyzed the effect of noisy feedback on the point-to-point MISO system and broadcast channel \cite{Housfater:09}. However, the authors did not incorporate the noisy feedback into the algorithm design.
In \cite{Jafarkhani:07, Tianyu:09}, the authors design a channel optimized quantizer for point-to-point MISO link to incorporate the noisy feedback link. However, extension to SDMA system is not-trivial. As illustrated in \cite{Hanzo:07}, the sensitivity of noisy feedback is much higher in SDMA systems and it is critical to take into consideration of noisy feedback in the robust limited feedback for SDMA systems.

When we have noisy CSIT feedback, there may be
significant performance degradation because an erroneous CSIT
feedback will make the base station selecting a wrong precoder for
the user, which not only decreases the received signal to noise
ratio (SNR) of the user but also increases the interference from
other scheduled users since their assigned precoders are no longer
orthogonal to the target user. As we shall illustrate in the paper,
adopting a {\em naive design approach} (design the limited feedback
codebook assuming error-free feedback and testing its performance in
the noisy limited feedback situation) will result in very poor SDMA
performance. In order to obtain a feedback-error-resilient design,
there are several first-order technical challenges to be addressed:
\begin{itemize}
\item{\bf{Robust SDMA precoding \& user selection:}} Due to the noisy limited
feedback, the CSIT index received at the base station may not be the
same as that sent by the mobiles. As such, the selected precoder may
not match the actual CSI at all, resulting in additional spatial
interference among the selected SDMA users.

\item{\bf{Robust rate adaptation:}} To achieve a high
system goodput advantage, robust rate adaptation is needed to
control packet errors due to channel outage.

\item{\bf{Robust Index Assignment:}} Index assignment refers to the mapping
of the CSIT feedback indices with the precoder entries in the
codebook. With noisy feedback, CSIT index assignment plays an
important role on the robustness performance of the SDMA systems.

\item{\bf{Performance Analysis:}} Beside robust limited feedback designs,
it is important to have closed-form performance results to obtain
useful design insights such as the sensitivity of CSIT errors in
SDMA systems.
\end{itemize}

In this paper, we propose a robust noisy limited feedback design
for SDMA systems. The solution consists of a real-time robust SDMA
precoding, user selection and rate adaptation as well as an offline
feedback index assignment algorithm. We formulate the robust index
assignment problem into a Traveling Salesman Problem (TSP) and
derive a low complex but asymptotically optimal solution. Simulation
results show that the proposed framework has significant goodput
gain compared to the traditional naive uncoded design (SDMA design
assuming error-free and uncoded feedback) and naive coded design
(SDMA design assuming error-free feedback but the limited CSIT
feedback bits are protected by FEC) under noisy limited feedback
channel. Furthermore, we show that despite the noisy feedback
channels, the average system goodput of the proposed robust SDMA
design scales as
$\mathcal{O}(\frac{n_T(1-\epsilon)}{n_T-1}(C_{fb}-\log_2(N_n)))$ for
interference limited scenario ($C_{fb} < (n_T-1)\log_2P+\log_2N_n$)
and $\mathcal{O}(n_T\cdot\log_2P)$ for noise limited scenario where
$n_T$ is the transmit antenna number, $\epsilon$ is the target outage probability,
$C_{fb}$ is the number of feedback bits, $P$ is the transmission power and $N_n$ is
a constant. We
find that in interference limited scenario, the average system
goodput increase linearly with the number of feedback bits for fixed
feedback SER and converge to a constant number for fixed feedback
SNR. In the noise limited scenario, the average system goodput
increase linearly with the number of transmit antenna and the
forward channel SNR ($\log_2P$).

\section{System Model}
\label{sect:sys-model}

In this paper, we shall adopt the following convention.
$\mathbf{X}$ denotes a matrix and $\mathbf{x}$ denotes a vector. $\mathbf{X}^\dag$
denotes matrix hermitian.

\subsection{Forward MIMO Fading Channel Model}
\label{sect:model-forward} In this paper, we consider a multi-user
system with a base station having $n_T$ transmit antennas
simultaneously transmitting to $n_T$ one antenna active users from a
total of K users. We shall focus on the case when $K>n_T$ so that
user scheduling in addition to precoder adaptation is important. The
base station separates $n_T$ data streams to the active users by
precoding. Each active user \emph{k} is assigned a $n_T\times 1$ precoding
vector $\mathbf{w}_k$. The precoder $\{\mathbf{w}_k\}_{k=1}^{n_T}$
are a set of unitary orthogonal vectors selected from a codebook of
multiple sets of unitary orthogonal vectors. Let $x_k$ be the
transmitted symbol of user \emph{k} with $E[|x_k|^2]=1$ and $y_k$ denote
the received symbol of user \emph{k}. The forward channel is modeled as:

\begin{equation}
y_k=\sqrt{\frac{P}{n_T}}\mathbf{h}_k^\dag\sum_{i\in
\mathcal{A}}\mathbf{w}_ix_i+z_k \label{eqn:forward-channel-model}
\end{equation}
where $P$ is the transmission power\footnote{In this paper we assume equal power allocation since power allocation will only bring marginal performance gain under high SNR. This is also assumed in \cite{Kaibin:07,Rao:05,Love:05} etc.}, $\mathbf{h}_k$ is the $n_T
\times 1$ complex channel state vector of the $k^{th}$ user,
$\mathcal{A}$ is the active user set and $z_k$ is the additive white
Gaussian noise with zero-mean and unit variance. We assume that the
transmit antennas and users are sufficient separated so that the
channel fading between different users and different antenna are
modeled as independent and identically distributed (i.i.d.) complex
Gaussian process with zero-mean and unit variance\footnote{In this
paper we assume the large scale fading parameters (path loss and shadowing)
between a base station and all the users in the cell are the same.}.

We consider slow fading channels where the fading is quasi-static
within a scheduling time slot for each user. This is a realistic
assumption for pedestrian mobility (~ 5km/hr) as the packet duration
is of the order of 500ns (such as Wi-Fi and B3G). Due to the quasi
static fading and noisy limited CSI feedback, there is uncertainty
on a user's instantaneous mutual information (a function of the
instantaneous CSI of all $K$ users) at the transmitter. Hence, there
exists potential packet errors (despite the use of powerful channel
coding) due to channel outage when the transmitted data rate of user
$\emph{k}$ exceeds its instantaneous mutual information.

\subsection{Limited Feedback Processing at the Mobiles}
\label{sect:Quantization-MS}

In this paper, we consider FDD system and assume the CSI is
estimated at each user (CSIR) perfectly and fed back to the base
station through a feedback channel with a limited feedback capacity
constraint $C_{fb}$ bits per fading block per user. The CSIR of user
\emph{k}, $\mathbf{h}_k$, consists of two parts: channel gain
$\|\mathbf{h}_k\|=\sqrt{\sum_{i=1}^{n_T} {h_k^i}^2}$ and channel
shape $\tilde{\mathbf{h}}_k=\mathbf{h}_k/\|\mathbf{h}_k\|$. As will
be explained in section \ref{sect:feedback-strategy}, the average
system goodput is dominated by channel shape and channel gain has
little influence on it especially in high SNR scenario\footnote{When there is only 1 active user in a cell, the system will degrade to a MISO system and this claim will be invalid. However, in MISO systems, interference is no longer a severe problem and there exists a lot of optimizing schemes. In this paper we will focus on strict SDMA system where there are more than 1 user served by the BS.}. Hence in
this paper, we shall focus on utilizing all the  $C_{fb}$
feedback bits on the feedback of channel shape
$\tilde{\mathbf{h}}_k$ and do not feed back channel gain
$\|\mathbf{h}_k\|$.

We assume the $K$ mobile stations quantize the channel shape of the
local CSIR with a codebook consists of multiple sets of orthonormal
vectors:

\begin{equation}
\mathcal{F}=\bigcup_{m=1}^M \mathcal{V}^{(m)} \label{eqn:codebook}
\end{equation}
where $\mathcal{F}$ denotes the quantization codebook with a
cardinality $|\mathcal{F}|=N$, $\mathcal{V}^{(m)}$ is the $m^{th}$
orthonormal set in the codebook and $M$ is the number of orthonormal
sets which is given by $M=\frac{N}{n_T}$. The $M$ orthonormal bases
of $\mathcal{F}$ are generated randomly and independently similar to
\cite{Zyczkowski:94}\footnote{Grassmannian codebook is not a good choice here. Maximizing the minimum distortion among M bases is not equivalent to maximizing the minimum distortion among all the vectors in the codebook. }. Define the distortion function between two
$n_T\times 1$ vectors $\mathbf{v}_1,\mathbf{v}_2$ as:

\begin{equation}
d(\mathbf{v}_1,\mathbf{v}_2)=1-|\mathbf{v}_1^\dag\mathbf{v}_2|^2=\sin^2(\angle(\mathbf{v}_1,\mathbf{v}_2))
\label{eqn:distortion}
\end{equation}

At the $\emph{k}-{th}$ mobile, the quantized channel shape
$\widehat{\widetilde{\mathbf{h}}}_k $ can be expressed as:

\begin{equation}
\widehat{\widetilde{\mathbf{h}}}_k = \arg\min_{\mathbf{v}\in
\mathcal{F}} d(\mathbf{v},\widetilde{\mathbf{h}}_k)
\label{eqn:quantized_h}
\end{equation}


In order to reduce the total feedback overhead for all $K$ users,
user \emph{k} shall decide whether to feed back its CSIT or not based on the
criteria:

\begin{equation}
d(\widetilde{\mathbf{h}}_k,\widehat{\widetilde{\mathbf{h}}}_k)<\delta;
\ \ \ \ \ |\mathbf{h}_k\|^2>g_{th} \label{eqn:ch-criteria}
\end{equation}
where $\delta$ is the threshold for the distortion from actual
channel shape to the quantized channel shape and $g_{th}$ is the
threshold for the channel gain. The motivation of
(\ref{eqn:ch-criteria}) is allocating the feedback bits to the users
with smaller CSI quantization error so as to reduce the potential
spatial interference among the SDMA streams. $\delta$ is a system
parameter and can be selected offline. When the number of total user
$K$ is large, $\delta$ can be selected as a very small number to
decrease the quantization error.

\subsection{CSIT Feedback Channel Model}
\label{sect:model-feedback}

Unlike most of the previous literature where the limited feedback
channel is assumed to be noiseless, we are interested in the more
realistic case where the feedback channel may be noisy. Note that
since the CSI feedback has to be delivered in a timely manner,
effective FEC coding over many CSI feedbacks is not possible and
hence, the feedback error cannot be ignored in practice.

The CSIT of user \emph{k} is quantized locally and encoded into $C_{fb}$
bits. The set of CSI indices sent at each of the K mobile stations
$\{\mathcal{I}^{(MS)}_k\}$ and the corresponding CSI indices
received at the base station $\{\mathcal{I}^{(BS)}_k\}$ both have
cardinality of $N=2^{C_{fb}}$. Assume the modulation symbol in the
CSIT feedback channel has constellation $\mathcal{M}$ and the
corresponding mapping from the CSIT index $\mathcal{I}^{(MS)}_k$ to
the constellation point $\mathbf{m}\in\mathcal{M}$ is given by the
1-1 {\em index mapping function}
$\mathcal{M}=\xi(\mathcal{I}^{(MS)}_k)$. The probabilistic
relationship between the CSIT feedback symbol sent
$\xi(\mathcal{I}^{(MS)}_k)$ and the CSIT feedback symbol received by
the transmitter $\xi(\mathcal{I}^{(BS)}_k)$ can be characterized by
the {\em feedback channel transition matrix}
$\mathbf{P}_{ch}=\{P_{m_l,m_k}^{ch}\}$, where:
\begin{eqnarray}
P_{m_l,m_k}^{ch}=\Pr\left[\xi(\mathcal{I}^{(BS)}_k)=m_l|\xi(\mathcal{I}^{(MS)}_k)=m_k\right],
 m_l,m_k\in\mathcal{M}. \label{eqn:transfer}
\end{eqnarray}
Hence, the {\em reliability} of the CSIT feedback channel is
characterized by the CSIT feedback channel transition matrix
$\mathbf{P}_{ch}$. Note that $\mathbf{P}_{ch}$ depends on the
average feedback SNR, feedback constellation and so on and can be offline
evaluated analytically or numerically through simulations.

Given $\mathbf{P}_{ch}$  and the index mapping rule $\xi$, the
stochastic relationship between the CSIT index sent by the mobile
station $\mathcal{I}^{(MS)}_k$ and the CSIT index received by the
base station $\mathcal{I}^{(BS)}_k$ is characterized by the {\em
CSIT index transition probability}
$\mathbf{P}_{CSIT}=\{P_{m_l,m_k}^{CSIT}\}$ given by:
\begin{equation}
P_{ij}^{CSIT}=
\Pr\left[\mathcal{I}^{(BS)}_k=j|\mathcal{I}^{(MS)}_k=i\right]=P_{\xi(i),\xi(j)}^{ch},
i,j\in[1,N]. \label{eqn:feedback-model2}
\end{equation}

Suppose the condition in (\ref{eqn:ch-criteria}) is satisfied for
the \emph{k}-th mobile, the index of $\widehat{\widetilde{\mathbf{h}}}_k$,
$\mathcal{I}^{(MS)}_k$, is then mapped to a constellation point
$m_k$ using an index mapping function
$m_k=\xi(\mathcal{I}^{(MS)}_k)$ and feed back to the base station.
Due to the noisy feedback channel, selection of the index mapping
function $\xi$ becomes important and will affect the robustness of
the SDMA system.

\section{Base Station Processing: SDMA Precoding, User Scheduling and Rate Adaptation} \label{sect:feedback-strategy}

In this section, we shall discuss the base station processing based
on the limited feedback sent from the K mobiles over a noisy
feedback channel with {\em index transition probability}
$\mathbf{P}_{CSIT}$. Specifically, we shall discuss the SDMA
precoding, user selection and rate adaptation.

\subsection{System Goodput}
\label{sect:system-goodput}

Consider a full multiplexing system where base station schedules
$n_T$ active users for SDMA. Define $\mathcal{A}$ as the active user
set and $\mathbf{w}_k$ as the precoder for a scheduled user \emph{k}. We
can write the instantaneous goodput (b/s/Hz successfully received by
user \emph{k}) $\rho_k$ for user $\emph{k}$ as:


\begin{equation}
\rho_k = r_k(\mathcal{I}^{(BS)}_k)\cdot 1[r_k(\mathcal{I}^{(BS)}_k)
< C_k(\mathbf{h}_k, \mathbf{w}_k)] \label{eqn:goodput-general}
\end{equation}
where $r_k$ is the data rate of the packet of user \emph{k} and is a
function of received CSIT at base station, $C_k$ is user $\emph{k}$'s
instantaneous mutual information and $1(A)$ is an indicator function
which is 1 if the event $A$ is true and 0 otherwise. User $\emph{k}$'s
instantaneous mutual information $C_k$ is a function of its channel
state $\mathbf{h}_k$ as well as the assigned precoder
$\mathbf{w}_k$. Specifically, $C_k$ can be written as:
\begin{equation}
C_k(\mathbf{h}_k, \mathbf{w}_k) =
\log_2(1+\frac{\frac{P}{n_T}|\mathbf{h}^\dag_k\mathbf{w}_k|^2}{1+\sum_{j\neq
k,j\in \mathcal{A}}\frac{P}{n_T}|\mathbf{h}^\dag_k\mathbf{w}_j|^2})
\label{eqn:inst-mutual}
\end{equation}

We define $\theta = \angle(\widetilde{\mathbf{h}}_k,\mathbf{w}_k)$
as the angle between the actual channel shape and the assigned
precoder for user \emph{k}. Similarly, define $\phi =
\angle(\widetilde{\mathbf{h}}_k,\mathbf{v}_{\mathcal{I}^{(MS)}_k})$
and $\varphi(\mathbf{v}_i,\mathbf{v}_j) =
\angle(\mathbf{v}_{i},\mathbf{v}_{j})$  as the angle between the
actual channel shape and the quantized vector of user \emph{k} and the
angle between two $n_T \times 1$ vectors $\mathbf{v}_i,
\mathbf{v}_j$.

Consider high effective SNR asymptotic scenario where
$\frac{P}{n_T}\|\mathbf{h}_k\|^2$ is sufficiently large. We can
further simplify (\ref{eqn:inst-mutual}) to:
\begin{eqnarray}
C_k &\approx& \log_2(1+\frac{\cos^2 \theta}{\sum_{j\neq k,j\in
\mathcal{A}}\cos^2\varphi(\widetilde{\mathbf{h}}_k,\mathbf{w}_j)})
\label{eqn:inst-mutual-high-SNR}
\end{eqnarray}

\begin{figure}
{\resizebox{9cm}{!}{\includegraphics{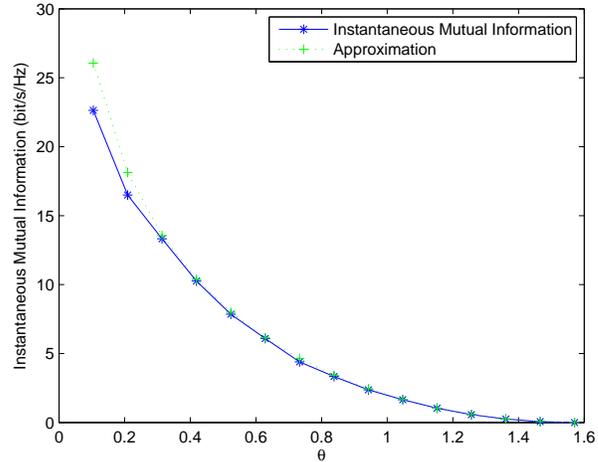}}}
\caption{Accuracy of approximation on instantaneous mutual
information at $SNR = 20dB$ and different $\theta$ in equation
\ref{eqn:inst-mutual-high-SNR}. } \label{fig:approx-capacity}
\end{figure}

As shown in Figure \ref{fig:approx-capacity}, the approximation is
quite good for moderate to high SNR.

Remark 1: While the approximation in (\ref{eqn:inst-mutual-high-SNR})
fails when $\sum_{j\neq k,j\in
\mathcal{A}}\cos^2\varphi(\widetilde{\mathbf{h}}_k,\mathbf{w}_j)\rightarrow
0$, it will not affect our design and analysis because for a
practical target PER (e.g. $10^{-2}$), when $\sum_{j\neq k,j\in
\mathcal{A}}\cos^2\varphi(\widetilde{\mathbf{h}}_k,\mathbf{w}_j)\rightarrow
0$, both $C_k$ in (\ref{eqn:inst-mutual}) and
(\ref{eqn:inst-mutual-high-SNR}) will be large enough and no outage
will occur. In other words, the case $\sum_{j\neq k,j\in
\mathcal{A}}\cos^2\varphi(\widetilde{\mathbf{h}}_k,\mathbf{w}_j)\rightarrow
0$ will not be the performance bottleneck and there is no loss of
generality to focus on the bottleneck case when $\sum_{j\neq k,j\in
\mathcal{A}}\cos^2\varphi(\widetilde{\mathbf{h}}_k,\mathbf{w}_j)$ is
not close to 0.

Remark 2: From equation (\ref{eqn:inst-mutual-high-SNR}), we can see that at high SNR, the instantaneous mutual information does not depend on the channel gain $\|\mathbf{h}_k\|$ and hence the average system goodput at high SNR is dominated by channel shape.

The average goodput for a scheduled user \emph{k} can be expressed as:
\begin{eqnarray}
\overline{\mathcal{\rho}}_k &=&
\mathbb{E}_{\mathcal{I}^{(BS)}_k}\mathbb{E}_{\mathbf{h}_k}[\mathcal{\rho}_k]\nonumber \\&=&
\mathbb{E}_{\mathcal{I}^{(BS)}_k}[r_k(\mathcal{I}^{(BS)}_k) \cdot
Pr(r_k<C_k|\mathcal{I}^{(BS)}_k)]
\label{eqn:avg-goodput}
\end{eqnarray}

In practice, there is a target PER requirement $\epsilon$ associated
with different application streams (e.g. $\epsilon = 0.01$ for voice
applications), which is expressed as:

\begin{equation}
1-Pr(r_k(\mathcal{I}^{(BS)}_k)<C_k|\mathcal{I}^{(BS)}_k) = \epsilon
\label{eqn:outage-constraint}
\end{equation}

The average system goodput for all scheduled users can be written
as:
\begin{eqnarray}
\mathcal{G}&=&\sum_{k\in \mathcal{A}}\overline{\rho}_k = n_T \cdot
(1-\epsilon) \cdot
\mathbb{E}_{\mathcal{I}^{(BS)}_k}[r_k(\mathcal{I}^{(BS)}_k)]
\label{eqn:avg-goodput-whole}
\end{eqnarray}

\subsection{SDMA Precoding and User Selection}
\label{sect:precoding-scheduling} Generally speaking, based on the
CSIT feedback, the active user set $\mathcal{A}$ and corresponding
precoder $\mathbf{w}_k$ shall be jointly optimized to maximize the
overall average system goodput in (\ref{eqn:avg-goodput-whole}).
Assume that the number of total user K is large enough such that the
base station can always fully schedule $n_T$ active users. A natural
algorithm includes an exhaustive search over all the combinations of
$n_T$ users out of a pool of K users and jointly optimize the $n_T$
precoders for the selected users to maximize the average system
goodput. However, exhaustive search has exponential complexity in
terms of number of users and is not practical as an online algorithm
especially for large number of users. Therefore we shall adopt a
simple orthogonal scheduling and precoding algorithm. At base
station, take the received CSIT $\mathbf{v}_{\mathcal{I}^{(BS)}_k}$
as the estimation of $\widetilde{\mathbf{h}}_k$, the term
$\cos^2\theta$ in the signal power term in
(\ref{eqn:inst-mutual-high-SNR}) is maximized if
$\mathbf{w}_k=\mathbf{v}_{\mathcal{I}^{(BS)}_k}$. Similarly, the
interference term $\sum_{j\neq k,j\in
\mathcal{A}}\cos^2\varphi(\widetilde{\mathbf{h}}_k,\mathbf{w}_j)$ is
minimized by choosing
$\mathbf{w}_j\perp\mathbf{v}_{\mathcal{I}^{(BS)}_k}, \ \ \forall\
j\neq k,j\in \mathcal{A}$.

It can be easily shown that the above two equations can be satisfied
simultaneously with:
\begin{equation}
\mathbf{v}_{\mathcal{I}^{(BS)}_k}\perp
\mathbf{v}_{\mathcal{I}^{(BS)}_j}, \ \ \forall j\neq k,
j,k\in\mathcal{A}. \label{eqn:schedule}
\end{equation}

In other words, the received CSIT of $n_T$ scheduled users
$\{\mathbf{v}_{\mathcal{I}^{(BS)}_k}\}$ form an orthonormal set
$\mathcal{V}^{(m)}$ in the codebook $\mathcal{F}$. In fact, SDMA
with orthogonal precoding is also called \emph{per unitary basis
stream user and rate control} (PU2RC) \cite{PU2RC} and has been
widely used in the standards such as 3GPP-LTE[3]. The main feature
of PU2RC is that it could accommodate limited CSIT feedback in a
natural way. For instance, the multiuser precoders are selected from
a codebook of multiple orthonormal bases. The importance of PU2RC
for the next-generation wireless communication motivates the
investigation of its performance in this paper. In this paper, we
consider a simplified PU2RC system where scheduled users have single
data streams, which are separated by orthogonal precoders.

The precoding and user scheduling strategy is briefly summarized as
follows:

\begin{itemize}
\item\emph{Step 1: search $N/n_T$ groups of orthonormal sets $\mathcal{V}^{(m)}$ in $\mathcal{F}$ for one set $\mathcal{V}^{(m)*}$ in which each vector is received by the base station as a CSIT feedback for at least one user. If there exists multiple satisfying sets, randomly pick one.}

\item\emph{Step 2: For an orthonormal vector $\mathbf{v}_k\in\mathcal{V}^{(m)*}$, randomly select one user from the group of users with received CSIT $\mathbf{v}_k$ as the scheduled user and set $\mathbf{v}_k$ as its precoder. }

\item\emph{Step 3: Repeat Step 2 for all the vectors in $\mathcal{V}^{(m)*}$}
\end{itemize}

\subsection{Robust Rate Adaptation}
\label{sect:rate-adaptation}
With the proposed precoding and user scheduling strategy, $C_k$ in
(\ref{eqn:inst-mutual-high-SNR}) is given by the following lemma:

\begin{Lem}[Mutual Information at High SNR]
At high downlink SNR, when the $n_T$ scheduled users are using $n_T$ orthonormal precoders to transmit, the mutual information of a scheduled user is given by:
\begin{equation}
C_k =
-2\log_2(\sin\theta). \label{eqn:inst-mutual-3}
\end{equation}
\label{lem:mutual_info}
\end{Lem}
\begin{Proof}
Please refer to Appendix A for details.
\end{Proof}

$r_k$ can be calculated from the requirement of the conditional PER
target in (\ref{eqn:outage-constraint}), which is given by:

\begin{equation}
P_{out} =
Pr(-2\log_2(\sin\theta)<r_k|\mathcal{I}_k^{(BS)}) = \epsilon
\label{eqn:rate-adaptation}
\end{equation}

Yet, one critical challenge in solving for $r_k$ in
(\ref{eqn:rate-adaptation}) is the knowledge of the CDF of
$\sin\theta$ conditioned on the CSIT $\mathcal{I}^{(BS)}_k$. This is
in contrast with the conventional approach of maximizing the ergodic
capacity in which only the first order moment of the random variable
$\sin\theta$ is needed. In the following lemma, we shall give a
tight upper bound on the conditional CDF of $\sin\theta$, which is
critical to solving for a closed-form rate adaptation solution.

\begin{Lem}[Upper bound of the PER $P_{out}$]
The conditional PER $P_{out}$ of the forward channel (conditioned on
the limited CSIT feedback received at the BS) is given by:
\begin{eqnarray}
P_{out}&\leq&
(1-\frac{(2^{-\frac{r_k}{2}}-\sin\varphi_{\mathcal{I}^{(BS)}_k,i^*})^{2(n_T-1)}}{\delta^{(n_T-1)}})
\cdot P^{CSIT}_{i^*,\mathcal{I}^{(BS)}_k}\nonumber \\& & + \sum_{j\in
\overline{Ns^{\epsilon}}_{\mathcal{I}_k^{(BS)}}}P^{CSIT}_{j,\mathcal{I}_k^{(BS)}}
\label{eqn:outage-upper-bound}
\end{eqnarray}
where $i^*= \arg\max_{i\in
Ns^{\epsilon}(\mathcal{I}^{(BS)}_k)}\sin\varphi_{\mathcal{I}^{(BS)}_k,i}$
and $Ns^{\epsilon}(\mathcal{I}^{(BS)}_k)$ is the set of neighboring
codewords of $\mathcal{I}^{(BS)}_k$ satisfying: $\sum_{j\in
Ns^{\epsilon}(\mathcal{I}^{(BS)}_k)}P^{CSIT}_{j,\mathcal{I}^{(BS)}_k}\geq
1-\epsilon$ and $\sum_{j\in
Ns^{\epsilon}(\mathcal{I}^{(BS)}_k)}P^{CSIT}_{j,\mathcal{I}^{(BS)}_k}-P^{CSIT}_{i^*,\mathcal{I}^{(BS)}_k}<
1-\epsilon.$ \label{lem:lemma1}
\end{Lem}
\begin{Proof}
Please refer to Appendix B for details.
\end{Proof}

Note that in practice, $i^*$ and  $Ns^{\epsilon}(\mathcal{I}^{(BS)}_k)$ can be offline precalculated from the channel
transition matrix $P^{CSIT}$ and the distortion between the codewords $\sin\varphi_{i,j}$.  For example, suppose our codebook is given by $\mathbf{F}=\{\mathbf{v_1,v_2,v_3,v_4}\}$. Assume $\epsilon = 0.1$ and $P^{CSIT}_{11}=0.70, \sin\angle(\mathbf{v_1,v_1})=0; P^{CSIT}_{21}=0.10,\sin\angle(\mathbf{v_1,v_2})=0.5; P^{CSIT}_{31}=0.11, \sin\angle(\mathbf{v_1,v_3})=0.4; P^{CSIT}_{41}=0.09, \sin\angle(\mathbf{v_1,v_4})=1$, then $Ns^{0.1}(1)=\{\mathbf{v_2,v_3}\}$ and $i^*=3$.

Using Lemma \ref{lem:lemma1}, and define
$\epsilon_{res}=\epsilon-\sum_{j\in
\overline{Ns^{\epsilon}}_{\mathcal{I}_k^{(BS)}}}P^{CSIT}_{j,\mathcal{I}_k^{(BS)}}$,
the transmission rate of user \emph{k} is given by:

\begin{equation}
r_k =
-2\log_2\left\{\delta^{\frac{1}{2}}(1-\frac{\epsilon_{res}}{P^{CSIT}_{i^*,\mathcal{I}^{(BS)}_k}})^{\frac{1}{2(n_T-1)}}+\sin\varphi_{\mathcal{I}^{(BS)}_k,i^*}\right\}.
\label{eqn:rk}
\end{equation}

\section{Robust Index Assignment}
\label{sect:index-assigment}

Index mapping algorithm is important when there is noise on feedback
channel. In this section, we shall optimize the index assignment
function $\xi$ to maximize the system goodput in
(\ref{eqn:avg-goodput-whole}). This is equivalent to minimize the
feedback distortion between the channel shape
$\widetilde{\mathbf{h}}_k$ and corresponding precoder of user \emph{k}
$\mathbf{v}_k^{(BS)}$, which is $\sin^2\theta$. This feedback
distortion is contributed by two parts: distortion from quantization
$\sin^2(\phi)$ and distortion from feedback error
$\sin^2(\varphi(\mathbf{v}_k^{(BS)},\mathbf{v}_k^{(MS)}))$. Based
on:
\begin{equation}
\varphi_{\mathcal{I}^{(BS)}_k,\mathcal{I}^{(MS)}_k}-\phi\leq\theta\leq
\varphi_{\mathcal{I}^{(BS)}_k,\mathcal{I}^{(MS)}_k}+\phi\label{eqn:theta-phi-varphi}
\end{equation}

and $\sin^2\phi<\delta$
where $\delta$ is chosen to be a small value to avoid excessive
spatial interference among the SDMA streams. As a result,we shall
omit quantization distortion $\sin^2\phi$ and focus on minimizing
the distortion introduced by feedback error
$d(\mathbf{v}_k^{(BS)},\mathbf{v}_k^{(MS)})=\sin^2(\varphi(\mathbf{v}_k^{(BS)},\mathbf{v}_k^{(MS)}))$.
The average distortion is given by:
\begin{equation}
\mathbb{E}(d)=\sum_{i=1}^N \sum_{j=1}^N Pr(\mathbf{v}_i) \cdot
P^{ch}_{\xi(i),\xi(j)} \cdot d(\mathbf{v}_i,\mathbf{v}_j)
\label{eqn:expected-distortion}
\end{equation}

Searching for optimal index mapping function $\xi$ can be summarized
into the following problem:

\begin{Problem}[Robust Index Assignment Problem]
Find an optimal index assignment function to minimize the average
distortion introduced by feedback error:
\begin{equation}
\xi^*=\arg\min_{\xi}\sum_{i=1}^N \sum_{j=1}^N Pr(\mathbf{v}_i) \cdot
P^{ch}_{\xi(i),\xi(j)} \cdot d(\mathbf{v}_i,\mathbf{v}_j)
\label{eqn:feedback-distortion}
\end{equation}
where $Pr(\mathbf{v}_i)$ is the probability that $\mathbf{v}_i$ is
the quantization output for $\mathbf{h}_k$ and
$d(\mathbf{v}_i,\mathbf{v}_j)$ is the distortion between two
codeword given in
(\ref{eqn:distortion}).\label{Prob:index-assignment}
\end{Problem}

Note that the insight of the above formulation is that a good index
mapping function shall map 2 precoders $\mathbf{v}_1,\mathbf{v}_{2}
\in \mathcal{F}$ with smaller distortion
$d(\mathbf{v}_1,\mathbf{v}_{2})$ to the constellation points
$m_1,m_2$ with larger transition probability $P^{ch}_{m_1,m_2}$.

In general, finding the optimal mapping $\xi(.)$ involves
combinatorial search. When the number of feedback bits is small, the
computation complexity of exhaustive search is still acceptable.
However, when the number of transmit and receive antennas gets
larger and more feedback bits are required, the exhaustive searching
time will increase double exponentially with $C_{fb}$. This
motivates the study on the low-complexity solution of the problem.

Consider a special case when the CSIT feedback index is modulated by
one $N$-PSK symbol. When feedback error occurs, the erroneous symbol
is likely to be one of the adjacent neighbors of the feedback
$N$-PSK symbol, which is referred as the nearest constellation
error. The average distortion introduced by feedback error in
(\ref{eqn:feedback-distortion}) can be simplified to:
\begin{eqnarray}
D(\xi)&=& P_e \cdot \{ (\frac{Pr(\mathbf{v}_{\xi^{-1}(1)})d(\mathbf{v}_{\xi^{-1}(1)},\mathbf{v}_{\xi^{-1}(2)})}{2}+\nonumber \\ & & \frac{Pr(\mathbf{v}_{\xi^{-1}(2)})d(\mathbf{v}_{\xi^{-1}(2)},\mathbf{v}_{\xi^{-1}(1)})}{2})+ \cdots +\nonumber \\
& &(\frac{Pr(\mathbf{v}_{\xi^{-1}(N)})d(\mathbf{v}_{\xi^{-1}(N)},\mathbf{v}_{\xi^{-1}(1)})}{2}+ \nonumber \\ & & \frac{Pr(\mathbf{v}_{\xi^{-1}(1)})d(\mathbf{v}_{\xi^{-1}(1)},\mathbf{v}_{\xi^{-1}(1)})}{2})\}
\label{eqn:distortion-adjacent-error}
\end{eqnarray}
where $P_e$ denotes the symbol error rate (SER) of the feedback
channel and $D(\xi)=\mathbb{E}(d)$ is the average distortion. Assume that $\{Nd_1, \cdots, Nd_N\}$ are $N$ virtual
cities, and the distance between the virtual cities $Nd_i$ and
$Nd_j$ is given by:
\begin{equation}
Dis(Nd_i,Nd_j)=P_e
\frac{Pr(\mathbf{v}_{i})d(\mathbf{v}_i,\mathbf{v}_j)+Pr(\mathbf{v}_{j})d(\mathbf{v}_j,\mathbf{v}_i)}{2}
\label{eqn:distance-CSIT}
\end{equation}

Equation~(\ref{eqn:distortion-adjacent-error}) can be expressed in
terms of distance between {\em virtual cities} as follows.
\begin{eqnarray}
D(\xi) &=&
Dis(Nd_{\xi^{-1}(N)},Nd_{\xi^{-1}(1)})+\nonumber \\
 & &Dis(Nd_{\xi^{-1}(1)},Nd_{\xi^{-1}(2)})+ \cdots +\nonumber \\
 & &Dis(Nd_{\xi^{-1}(N-1)},Nd_{\xi^{-1}(N)})
\label{eqn:distance-TSP}
\end{eqnarray}

Hence, the optimization metric is equivalent to the total distance
of a Hamiltonian cycle \cite{Jensen:00}. From equation
(\ref{eqn:distance-TSP}), the index mapping problem in
Problem~\ref{Prob:index-assignment} is equivalent to searching
shortest path in a Hamiltonian cycle and this can be cast into a
{\em traveling salesman problem} (TSP). This is summarized below.

\begin{Problem}[Traveling Salesman Problem]
Given a number of cities $\{Nd_1, Nd_2, \cdots, Nd_N\}$,  and the
costs of traveling from any city to any other city $\{Dis( Nd_i,
Nd_j)\}$, what is the round-trip route
$[\xi^{-1}(1),\xi^{-1}(2),\cdots,\xi^{-1}(N)]$  that visits each
city exactly once and then returns to the starting city to minimize
the total distance (\ref{eqn:distance-TSP}). \label{Prob:tsp}
\end{Problem}

TSP is found to be an NP-hard (nondeterministic polynomial time)
problem and yet, there are a number of efficient searching
algorithms for the TSP such as the cutting-plane method
\cite{David:01} and genetic algorithm\cite{PL:99}.

In this paper, we propose a simple construction algorithm, namely
the \emph{Circled Nearest Neighbor Algorithm}(CNNA). The CNNA
algorithm is described below:

\begin{itemize}
    \item\emph{Step 1: Start the TSP travel from a randomly selected node $Nd_i$.}

    \item\emph{Step 2: Go to the nearest unvisited node from $Nd_i$. If there exists more than one such nodes, we select the one with smallest sum distortion to the previously visited nodes.
    For example, if $\mathcal{S}$ is the set of nodes already visited, and $\mathcal{N}_i$ is the set of unvisited neighboring nodes of $Nd_i$, we shall select $Nd_j$ via the criteria: $Nd_j=\arg\min_{Nd_j\in\mathcal{N}_i}\sum_{Nd_k\in\mathcal{S}}dis(Nd_k,Nd_j)$}

    \item\emph{Step 3: Repeat Step 2 till all the nodes are visited. Then go back to the start node.}
\end{itemize}

\begin{figure}
{\resizebox{9cm}{!}{\includegraphics{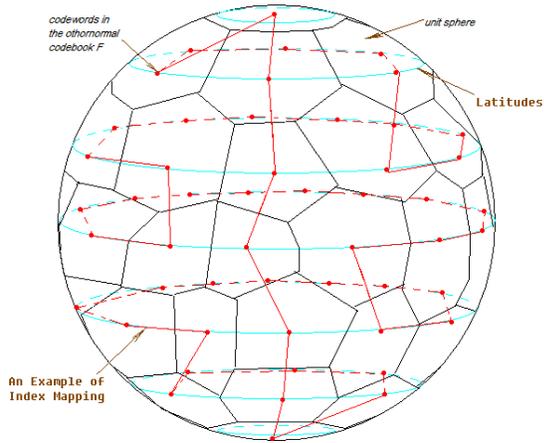}}}
\caption{A heuristic TSP travel on a hypersphere.}
\label{fig:TSP-ball}
\end{figure}

Suppose we require that all feedback error results in the "nearest
neighboring precoder" and we would like to find an index assignment
such that this can be realized for "2-nearest neighbor error
feedback channel". To do that, we assume all the neighboring regions
of codewords are "equi-probable" and isotropically distributed on
the surface of a unitary hypersphere
($\widetilde{\mathbf{h}}_k$-space). For the given topology of the
partition region, the above algorithm of index assignment will
result in a circled pattern as illustrated in figure
\ref{fig:TSP-ball} and is similar to the following analogy. Suppose
we start from north pole of the Earth and travel around the world
along the latitudes to the south pole. When we finish traveling
along a latitude, we go down to the next one until we reach the
south pole. After south pole is arrived, we go back to the north
pole directly. Define $d_{min}$ as the minimum distortion between
two precoder, we have the following lemma:

\begin{Lem}[Asymptotic Optimality of CNNA Algorithm] For N-PSK constellation
with nearest-constellation error approximation, the index mapping
solution given by the CNNA algorithm $\xi^*$ is asymptotically
optimal for sufficiently large N. i.e.

\begin{equation}
\lim_{N\rightarrow +\infty}\frac{D(\xi^*)}{N}=d_{min}.
\label{eqn:index-optimal}
\end{equation}
\label{lem:index-assignment-optimal}
\end{Lem}
\begin{Proof} We provide a sketch of proof due to page limit.
With the proposed algorithm, all error events shall
result in nearest neighbor codeword errors (for 2-nearest
constellation feedback channel) except for the codewords serving as
the starting and ending nodes. Hence we have
\begin{equation}
D=N\cdot d_{min} + c \label{eqn:D-optimal}
\end{equation}
where $c$ is a constant.  When $N$ is sufficiently large, the
average distortion of the travel is $\lim_{N\rightarrow
+\infty}\frac{D}{N}=d_{min}$. Hence the proposed algorithm is
asymptotically optimal.
\end{Proof}

\section{Performance Analysis}
\label{sect:performance-analysis}

In this section, we shall focus on obtaining the asymptotic goodput
performance under noisy limited feedback. The transmission rate for
a scheduled user in (\ref{eqn:rk}) can be bounded by:
$-2\log_2(\delta^\frac{1}{2}+\sin\varphi_{\mathcal{I}^{(BS)}_k,i^*})
\leq r_k \leq -2\log_2(\sin\varphi_{\mathcal{I}^{(BS)}_k,i^*})$.

In fact, the upper bound and lower bound of $r_k$ above are both
very tight when $\delta$ is small. Asymptotically, the two bounds
will meet each other as the number of feedback bits $C_{fb}$ goes to
infinity since $\delta$ will approach to 0 and
$\sin\varphi_{\mathcal{I}^{(BS)}_k,i^*}$ is the dominant factor to
$r_k$. Since we are interested in the first-order analysis, the data
rate $r_k$ can be taken as:

\begin{equation}
r_k = \mathcal{O}(-2\log_2(\sin\varphi_{\mathcal{I}^{(BS)}_k,i^*}))
\label{eqn:rk-approx}
\end{equation}

Substitute (\ref{eqn:rk-approx}) into (\ref{eqn:avg-goodput-whole}),
the average system goodput is given by:
\begin{eqnarray}
\mathcal{G}&=&
\mathcal{O}(-2n_T(1-\epsilon)\log_2(\mathbb{E}_{\mathcal{I}^{(BS)}_k}[\sin\varphi_{\mathcal{I}^{(BS)}_k,i^*}]))
\label{eqn:avg-goodput-2}
\end{eqnarray}

From (\ref{eqn:avg-goodput-2}), we can see that system goodput
performance depends on the worst case distortion
$\sin\varphi_{\mathcal{I}^{(BS)}_k,i^*}$ of an index assignment
function $\xi$. Define $N_n^\epsilon(i)$ as the set of neighboring
points of a constellation point $i$ (including itself) with
$\sum_{j\in N_n^\epsilon(i)} P_{ch}(i,j)=1-\epsilon$. The
cardinality of the set is $N_n=|N_n^\epsilon(i)|$. Note that
$N_n^\epsilon(i)$ is the $N_n$ largest terms in the $i_th$ row of
$P_{ch}$.\footnote{Example: For 8PSK with 10dB feedback SNR we have
$N_n=3$ when $\epsilon=0.03$.} As a result, both $P_e$ and $N_n$ are
two first order parameters to characterize the quality of feedback
channel. We have the following lemma:

\begin{Lem}[Lower bound for $\mathbb{E}_{\mathcal{I}^{(BS)}_k}(\sin\varphi_{\mathcal{I}^{(BS)}_k,i^*})$]
For sufficiently large N, we have:
\begin{equation}
\mathbb{E}_{\mathcal{I}^{(BS)}_k}(\sin\varphi_{\mathcal{I}^{(BS)}_k,i^*})\geq
(\frac{N_n}{N})^{\frac{1}{2(n_T-1)}} \label{eqn:istar-distortion}
\end{equation}
\label{lem:lemma2}
\end{Lem}
\begin{Proof}
Please refer to Appendix C for details.
\end{Proof}

\begin{figure}
{\resizebox{9cm}{!}{\includegraphics{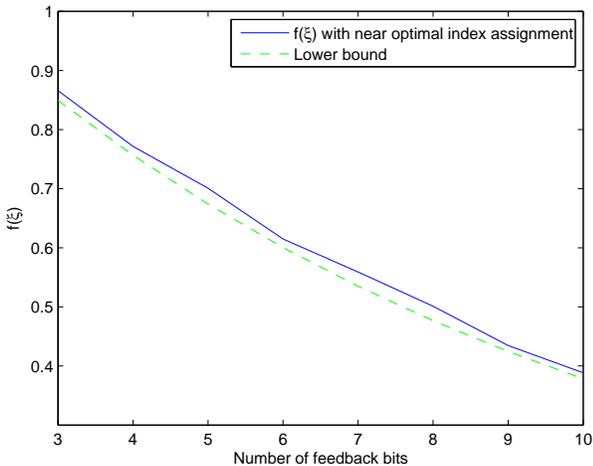}}}
\caption{Accuracy of approximation on
$f(\xi)=\mathbb{E}_{\mathcal{I}^{(BS)}_k}(\sin\varphi_{\mathcal{I}^{(BS)}_k,i^*})$
in Lemma \ref{lem:lemma2}. } \label{fig:approx-TSP}
\end{figure}

Numerical results show that with optimal or near optimal index
assignment, the lower bound is quite tight as illustrated in figure
\ref{fig:approx-TSP}. Take the lower bound in
(\ref{eqn:istar-distortion}) as an approximation of
$\mathbb{E}_{\mathcal{I}^{(BS)}_k}(\sin\varphi_{\mathcal{I}^{(BS)}_k,i^*})$
and substitute into (\ref{eqn:avg-goodput-2}), we have the following
theorem:

\begin{Thm}[Asymptotic System Goodput $\mathcal{G}$ in Interference Dominant Scenario]
When the number of feedback bits $C_{fb}$ satisfies
$C_{fb}<(n_T-1)\log_2P+\log_2N_n$, the system is dominated by
interference. For sufficiently large transmission power P and
quantization codebook size N, the average system goodput is given
by:
\begin{eqnarray}
\mathcal{G}&=&
\mathcal{O}(\frac{n_T(1-\epsilon)}{n_T-1}(C_{fb}-\log_2(N_n)))
\label{eqn:avg-goodput-inter-limited}
\end{eqnarray}
\label{lem:theorem-1}
\end{Thm}

\begin{Cor}
In interference dominant system with sufficiently large transmission
power P and quantization codebook size N:

\begin{itemize}
\item{With fixed SER $P_e$ on the feedback channel, $N_n$ is a finite constant and the average system
goodput is given by:}
\begin{eqnarray}
\mathcal{G}&=& \mathcal{O}(\frac{n_T(1-\epsilon)}{n_T-1}C_{fb})
\label{eqn:avg-goodput-inter-limited-fix-SER}
\end{eqnarray}
\item{With fixed feedback SNR, $N_n$ scales with $N$ as $N_n=c\cdot2^{C_{fb}}$ where c is a
constant and the average system goodput is given by:}
\begin{eqnarray}
\mathcal{G}&=& \mathcal{O}(\frac{n_T(1-\epsilon)}{n_T-1})
\label{eqn:avg-goodput-inter-limited-fix-SNR}
\end{eqnarray}
\end{itemize}
\label{cor:cor1}
\end{Cor}

\begin{figure}
{\resizebox{9cm}{!}{\includegraphics{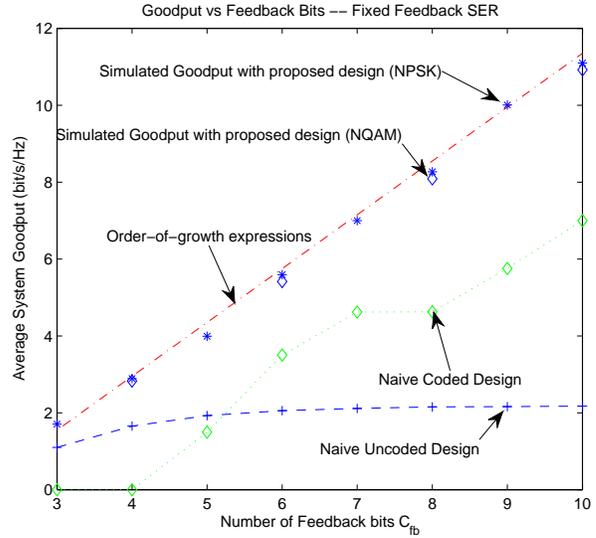}}}
\caption{System goodput versus number of feedback bits. Naive
Uncoded Design refers to design assuming the limited feedback
channel is noiseless and the feedback bits are uncoded and Naive
Coded Design is similar to naive uncoded design except the feedback
bits are protected by Hamming code. The SER on feedback channel is
fixed at 0.2. The dotted curve of naive coded design is not smooth
because of the overhead in the parity bits of Hamming code $(k,2^k)$
used to protect the CSIT feedback. } \label{fig:goodput-cfb-fxser}
\end{figure}

\begin{figure}
{\resizebox{9cm}{!}{\includegraphics{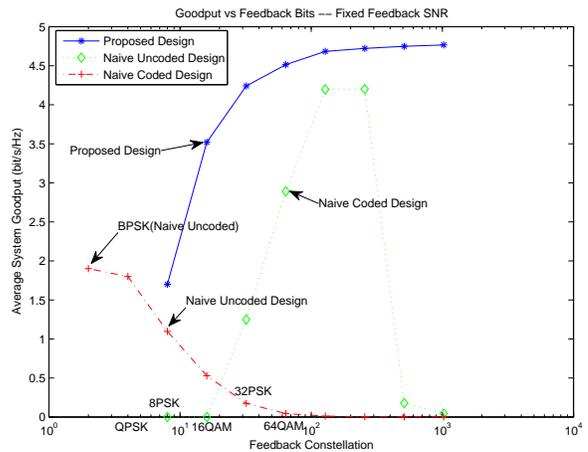}}}
\caption{Average system goodput versus Limited Feedback constellation Level. Naive
Uncoded Design refers to the limited feedback design (assuming noiseless feedback) and the feedback bits are uncoded and Naive
Coded Design is similar to naive uncoded design except the feedback
bits are protected by Hamming code. In all the schemes, the feedback
SNR(10dB) and the number of feedback symbols are fixed. The dotted curve of naive coded design drops
dramatically when the number of feedback bits gets large because
Hamming code only protects nearest neighbor errors. }
\label{fig:goodput-cfb-fxsnr}
\end{figure}

The goodput order of growth results in
(\ref{eqn:avg-goodput-inter-limited-fix-SER}) and
(\ref{eqn:avg-goodput-inter-limited-fix-SNR}) are also verified
against simulations in Figure \ref{fig:goodput-cfb-fxser} and Figure
\ref{fig:goodput-cfb-fxsnr}.

On the other hand, when the number of feedback bits is sufficiently
large, the SDMA system will operate in the noise-dominated regime.
\begin{Thm}[Asymptotic System Goodput $\mathcal{G}$ in Noise Dominant Scenario]
When the number of feedback bits $C_{fb}$ satisfies
$C_{fb}>(n_T-1)\log_2P+\log_2N_n$, the system is noise-dominated.
For sufficiently large transmission power P and quantization
codebook size N, the system goodput is given by:
\begin{eqnarray}
\mathcal{G}&=&
\mathcal{O}(n_T\cdot\log_2P)\label{eqn:avg-goodput-noise-limited}
\end{eqnarray}
\label{lem:theorem-2}
\end{Thm}
\begin{Proof}
Please refer to Appendix D for details.
\end{Proof}

The above result also reduces to that for SDMA system with noiseless
feedback when $N_n=\mathcal{O}(1)$ . Note that $N_n =
\mathcal{O}(1)$ actually corresponds to an asymptotically noiseless
feedback channel. We shall note that
$C_{fb}>(n_T-1)\log_2P+\log_2N_n$ may not be able to be satisfied by
simply increasing $C_{fb}$ since $C_{fb}$ will be canceled when
$N_n=c\cdot2^{C_{fb}}$ (e.g. constant feedback SNR). This indicates
that one could not enhance the CSIT quality by increasing $C_{fb}$
if the feedback SNR is kept constant.

\section{Results and Discussions}
\label{sect:results} In this section, we study the performance of
the proposed robust SDMA system under noisy limited feedback. We
compare the performance of proposed design with two naive designs
under the same feedback cost. In the uncoded naive design, the SDMA
system is designed as if the limited CSIT feedback were noiseless
and the limited CSIT feedback bits are uncoded. In the coded naive
design, the SDMA system is similar to the uncoded naive design
except that the limited CSIT feedback bits are protected by hamming
code. In the simulations, we consider an SDMA system with $n_T = 4$
forward SNR 20dB and $K=100$. We set the thresholds $\delta=0.1$ and $g_{th}=2$.

\subsection{System Performance with respect to the Number of Feedback Bits $C_{fb}$}
\label{sect:results-cfb}

Figure \ref{fig:goodput-cfb-fxser} illustrates the average goodput
versus the number of feedback bits of the SDMA system with a fixed
feedback SER of 0.2\footnote{This corresponds to a feedback SNR of
10 dB when $C_{fb}=3$ and 8PSK is adopted.}. There are significant
goodput gain compared with both the naive uncoded and coded designs.
The order-of-growth expressions in Theorem \ref{lem:theorem-1} and
Corollary \ref{cor:cor1} are also verified.

Figure \ref{fig:goodput-cfb-fxsnr} illustrates the average goodput
versus the modulation level per feedback symbol with a fixed average
feedback SNR of 10dB and a fixed number of feedback symbols. There are also significant goodput gain in the proposed scheme
compared with both the naive uncoded and coded designs. In addition, there is a tradeoff relationship in the feedback constellation level for the baseline systems. With lower feedback constellation level (such as BPSK), the feedback is more robust but the average goodput performance is limited by the resolution in the CSI feedback. On the other hand, for large feedback constellation, the average goodput performance of the reference baselines are poor because the performance is limited by the feedback error.

\subsection{System Performance with respect to Feedback Quality}
\label{sect:results-fb-quality}

\begin{figure}
{\resizebox{9cm}{!}{\includegraphics{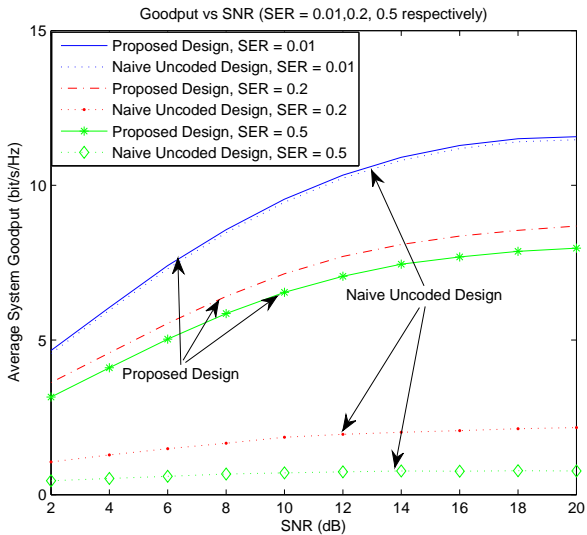}}}
\caption{System goodput versus Forward SNR at different level of
feedback SER. Number of feedback bits $C_{fb}=8$. With proposed
design, the system goodput decreases much slower with the increasing
of feedback SER.} \label{fig:goodput-ser}
\end{figure}

\begin{figure}
{\resizebox{9cm}{!}{\includegraphics{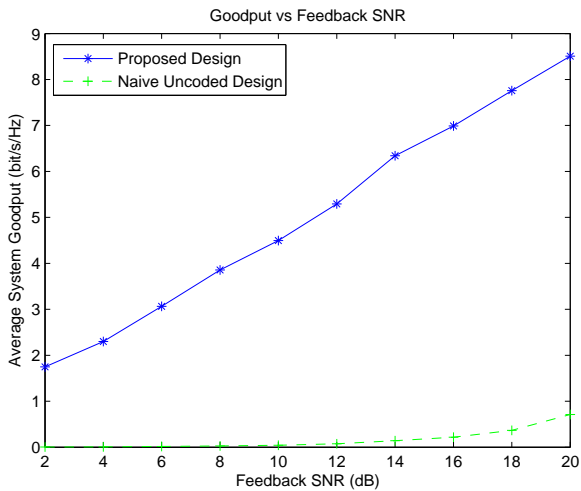}}}
\caption{System goodput versus feedback SNR. Number of feedback bits
$C_{fb}=6$ and forward transmission SNR is 20 dB. The performance of
the naive design is very bad since the $N_n$ precoders that may be
adopted when feedback error occur are randomly selected and the
transmission rate is determined by the worst precoder.}
\label{fig:goodput-snr}
\end{figure}

The feedback quality can be specified by feedback SER and feedback
SNR. Figure \ref{fig:goodput-ser} shows the average system goodput
versus SNR with different feedback SER and fixed feedback bits
$C_{fb}=8$. It is shown that with the proposed design, the system
goodput decrease much slower with the increasing of the feedback
SER. In figure \ref{fig:goodput-snr}, we show the average system
goodput versus feedback SNR with fixed feedback bits $C_{fb}=6$.
With proposed design the system goodput increases much faster with
the increasing of the feedback SNR.

\section{Conclusion}
\label{sect:summary} In this paper, we proposed a robust noisy
limited feedback design with a joint user scheduling and precoder
scheme as well as rate adaptation and robust index assignment
optimization algorithms. We convert the index assignment
optimization problem to a \emph{Traveling Salesman Problem}(TSP).
Simulation results show that the proposed framework has significant
goodput gain compared to the uncoded and coded naive designs.
Furthermore, we show that despite the noisy feedback, the average
system goodput scales as
$\mathcal{O}(\frac{n_T(1-\epsilon)}{n_T-1}(C_{fb}-\log_2(N_n)))$ and
$\mathcal{O}(n_T\cdot\log_2P)$ in the interference limited regime
($C_{fb} < (n_T-1)\log_2P+\log_2N_n$) and noise-limited regime
respectively.

\section*{Appendix-A: Proof of Lemma \ref{lem:mutual_info}}
From equation (\ref{eqn:inst-mutual}), the instantaneous mutual information can be
written as:
\begin{eqnarray}
C_k(\mathbf{h}_k, \mathbf{w}_k)
&\approx& \log_2(1+\frac{\cos^2 \theta}{\sum_{j\neq k,j\in
\mathcal{A}}\cos^2\varphi(\widetilde{\mathbf{h}}_k,\mathbf{w}_j)})
\nonumber \\ &=&-2\log_2(\sin\theta)
\label{eqn:inst-mutual-deduction}
\end{eqnarray}

(\ref{eqn:inst-mutual-deduction}) is because all the scheduled precoders $\mathbf{w}_j$ forms an orthogonal bases of the $n_T$ dimensional space and
$(\cos{\varphi(\widetilde{\mathbf{h}}_k,\mathbf{w}_{j1})},\cdots,\cos{\varphi(\widetilde{\mathbf{h}}_k,\mathbf{w}_{jn_T})})$  is a vector on the unit sphere of the $n_T$ dimensional space. where $\mathbf{w}_{j1}$ to $\mathbf{w}_{jn_T}$ are $n_T$ scheduled orthonomal precoders. Hence we have:

\begin{equation}
\cos^2\theta + \sum_{j\neq k, j\in
\mathcal{A}}\cos^2{\varphi(\widetilde{\mathbf{h}}_k,\mathbf{w}_{j})})=1.
\label{eqn:theta-relation}
\end{equation}

\section*{Appendix-B: Proof of Lemma \ref{lem:lemma1}} Given
received CSIT $\mathcal{I}^{(BS)}_k$, the outage probability can be
written as:
\begin{eqnarray}
P_{out}&\approx&
\frac{Pr(2^{-\frac{r_k}{2}}-\sin\varphi_{\mathcal{I}^{(BS)}_k,i^*}<\sin\phi\leq
\sqrt{\delta})}{Pr(\sin\phi<\sqrt{\delta})} \cdot
P^{CSIT}_{i^*,\mathcal{I}^{(BS)}_k} \nonumber \\ & &+ \sum_{j\in
\overline{Ns^{\epsilon}}_{\mathcal{I}_k^{(BS)}}}P^{CSIT}_{j,\mathcal{I}_k^{(BS)}}
\label{eqn:outage-upper-bound-append}
\end{eqnarray}
where $i^*= \arg\max_{i\in
Ns^{\epsilon}(\mathcal{I}^{(BS)}_k)}\sin\varphi_{\mathcal{I}^{(BS)}_k,i}$
and $Ns^{\epsilon}(\mathcal{I}^{(BS)}_k)$ is the set of neighboring
codewords of $\mathcal{I}^{(BS)}_k$ satisfying: $\sum_{j\in
Ns^{\epsilon}(\mathcal{I}^{(BS)}_k)}P^{CSIT}_{j,\mathcal{I}^{(BS)}_k}\geq
1-\epsilon$ and $\sum_{j\in
Ns^{\epsilon}(\mathcal{I}^{(BS)}_k)}P^{CSIT}_{j,\mathcal{I}^{(BS)}_k}-P^{CSIT}_{i^*,\mathcal{I}^{(BS)}_k}<
1-\epsilon$. From \cite{Dai:08}, we have the CDF for $\sin\phi$:
\begin{equation}
Pr(\sin\phi<x) = x^{2(n_T-1)} \label{eqn:distribution-sinphi}
\end{equation}
Subscribing (\ref{eqn:distribution-sinphi}) into
(\ref{eqn:outage-upper-bound-append}), we can simplify the outage
probability as:

\begin{eqnarray}
P_{out}&\leq&
(1-\frac{(2^{-\frac{r_k}{2}}-\sin\varphi_{\mathcal{I}^{(BS)}_k,i^*})^{2(n_T-1)}}{\delta^{(n_T-1)}})
\cdot P^{CSIT}_{i^*,\mathcal{I}^{(BS)}_k} \nonumber \\ & &+ \sum_{j\in
\overline{Ns^{\epsilon}}_{\mathcal{I}_k^{(BS)}}}P^{CSIT}_{j,\mathcal{I}_k^{(BS)}}.
\nonumber
\end{eqnarray}

\section*{Appendix-C: Proof of Lemma \ref{lem:lemma2}}

The asymptotic expression for
$\mathbb{E}_{\mathcal{I}^{(BS)}_k}(\sin\varphi_{\mathcal{I}^{(BS)}_k,i^*})$
is deduced assuming all the channel shape quantized to the elements
in $Ns^{\epsilon}(\mathcal{I}_k^{(BS)})$ forms a neighboring region
of $\mathbf{v}_{\mathcal{I}_k^{(BS)}}$ defined as a hypersphere:

\begin{equation}
\mathcal{C}_b(\mathbf{v}_{\mathcal{I}_k^{(BS)}})=\{\widetilde{\mathbf{h}}_k:\sin\angle(\widetilde{\mathbf{h}}_k,\mathbf{v}_{\mathcal{I}_k^{(BS)}})\leq
r_{\mathcal{C}_b}\} \label{eqn:partition-ball}
\end{equation}
where $r_{\mathcal{C}_b}$ is the radius of the hypersphere and
$\mathbf{v}_{i^*}$ lies at the edge of the hypersphere. Since
$Pr(\sin\angle(\widetilde{\mathbf{h}}_k,\mathbf{v}_i)<x)=x^{2(n_T-1)}$,
and the size of the hypersphere is:
\begin{equation}
Pr(\sin\angle(\widetilde{\mathbf{h}}_k,\mathbf{v}_{\mathcal{I}_k^{(BS)}})\leq
r_{\mathcal{C}_b}) =
\frac{|Ns^{\epsilon}(\mathcal{I}_k^{(BS)})|}{N}\geq \frac{N_n}{N}
\label{eqn:size-ball}
\end{equation}
where $|Ns^{\epsilon}(\mathcal{I}_k^{(BS)})|$ is the cardinality of
$Ns^{\epsilon}(\mathcal{I}_k^{(BS)})$. Therefore we have $r_{\mathcal{C}_b} \geq
(\frac{N_n}{N})^{\frac{1}{2(n_T-1)}}$


Since $\mathbf{v}_{i^*}$ lies at the edge of the hypersphere, we
have:
\begin{equation}
\mathbb{E}_{\mathcal{I}^{(BS)}_k}(\sin\varphi_{\mathcal{I}^{(BS)}_k,i^*})\geq
(\frac{N_n}{N})^{\frac{1}{2(n_T-1)}}.\nonumber
\end{equation}

\section*{Appendix-D: Proof of Theorem \ref{lem:theorem-2} }

Consider the situation with noise power larger than interference
power $1>\mathbb{E}[\frac{P}{n_T}\|\mathbf{h}_k\|^2\sin^2\theta]$ as
noise dominant scenario. With optimal or near optimal index
assignment, we have $\mathbb{E}[\sin^2\theta]\leq
(\frac{N_n}{N})^{\frac{1}{n_T-1}}$. Therefore, we require
$P\cdot(\frac{N_n}{N})^{\frac{1}{n_T-1}}<1$ which gives
$C_{fb}>(n_T-1)\log_2P+\log_2N_n $.
The instantaneous mutual information of a user in
(\ref{eqn:inst-mutual}) can be simplified into
$C_k \approx \log_2(1+\frac{P}{n_T}\|h_k\|^2\cos^2\theta)$.
As the channel gain $\|\mathbf{h}_k\|^2$ scale with $n_T$, without
loss of generality, we can also select $g_{th}$ in
(\ref{eqn:ch-criteria}) at order $O(n_T)$. Therefore the
transmission rate for user \emph{k} has the order $r_k \sim O(\log_2n_T)$
hence the system goodput $\mathcal{G}$ has the order:

\begin{eqnarray}
\mathcal{G}&\sim&
O(n_T\cdot\log_2P).\label{eqn:avg-goodput-noise-limited}
\end{eqnarray}

\begin{biography}[{\includegraphics[width=1in,height=1.25in,clip,keepaspectratio]{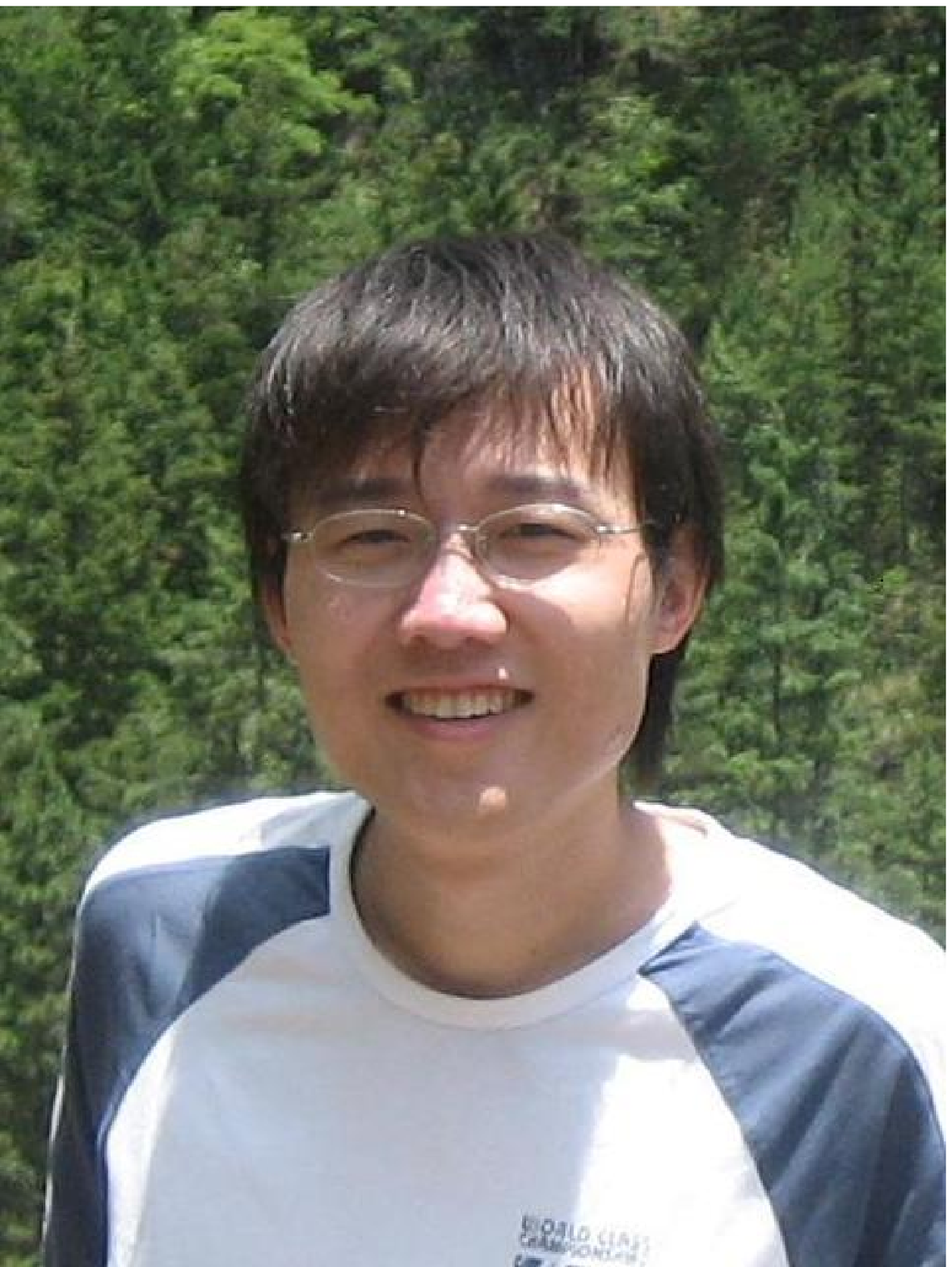}}]{T.Y. Wu}
received B.Eng degree from Tsinghua University in 2001 and Ph.D degree from Hong Kong
University of Science and Technology in 2009 respectively. He is currently a research
engineer in Huawei Technologies Co.Ltd. His research interests include limited feedback, interference
management,  and cognitive radio networks.
\end{biography}

\begin{biography}[{\includegraphics[width=1in,height=1.25in,clip,keepaspectratio]{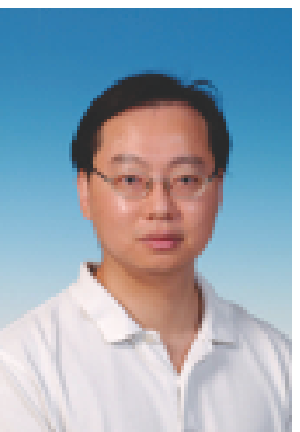}}]{Vincent K.N. Lau}
obtained B.Eng (Distinction 1st Hons) from the University of Hong
Kong (1989-1992) and Ph.D. from the Cambridge University
(1995-1997). He joined the Bell Labs - Lucent Technologies as member
of technical staff from 1997-2003 and the Department of ECE, Hong
Kong University of Science and Technology (HKUST) as Associate
Professor afterwards. His current research focus includes robust
cross layer scheduling for MIMO/OFDM wireless systems with imperfect
channel state information, communication theory with limited
feedback as well as delay-sensitive cross layer optimizations. He is
currently an associate editor of IEEE Transactions on Wireless
Communications, IEEE JSAC, EUARSIP Wireless Communications and
Networking.
\end{biography}




\end{document}